**Preparation, characterization, and electrical properties of epitaxial $NbO_2$ thin film lateral devices**

Toyanath Joshi, Tess R. Senty, Pavel Borisov, Alan D. Bristow and David Lederman

*Department of Physics and Astronomy, West Virginia University, Morgantown, WV 26506-6315, USA*

**Abstract**

Epitaxial $NbO_2$ (110) films, 20 nm thick, were grown by pulsed laser deposition on $Al_2O_3$ (0001) substrates. The $Ar/O_2$ total pressure during growth was varied to demonstrate the gradual transformation between $NbO_2$ and $Nb_2O_5$ phases, which was verified using x-ray diffraction, x-ray photoelectron spectroscopy, and optical absorption measurements. Electric resistance threshold switching characteristics were studied in a lateral geometry using interdigitated Pt top electrodes in order to preserve the epitaxial crystalline quality of the films. Volatile and reversible transitions between high and low resistance states were observed in epitaxial $NbO_2$ films, while irreversible transitions were found in case of $Nb_2O_5$ phase. Electric field pulsed current measurements confirmed thermally-induced threshold switching.

## 1. Introduction

Recently there has been a growing interest in materials demonstrating metal-insulator transitions (MITs) because of their possible applications in electronic devices [1-15]. Among these materials is $NbO_2$ which exhibits one of the highest MIT temperatures of 1081K [16-17], accompanied by a structural transition from a distorted rutile (low temperature) to a rutile structure (high temperature phase). The low temperature $NbO_2$ has a tetragonal unit cell (space group $I4_1/a$, $a_T$ = 13.702 Å and $c_T$ = 5.985 Å [16]), whereas the high temperature phase is described by a rutile unit cell (space group $P4_2/mnm$, a = 4.846 Å and c = 3.032 Å [16]).

Reversible threshold resistance switching has been reported in $NbO_2$ thin film devices at room temperature caused by the local heating effect triggering the MIT [1, 2]. The current-voltage characteristics exhibit current-controlled (S-type) negative resistance while switching from a low current semiconducting to a high current metallic state [1, 18-20]. Variations of the Nb oxidation state in polycrystalline $NbO_x$ films have also been reported after an electroforming current pulse was sent [21]. For example, the insulating $Nb_2O_5$ phase can be reduced to metallic NbO and semiconducting $NbO_2$, the latter being the predominant phase [21], so that conducting filaments are created or destroyed, respectively. As a result, non-volatile resistive switching of the thin film element is possible [3, 4]. Thus, $NbO_x$ thin films could be used for memory devices and electrical switching applications, while stoichiometric $NbO_2$ junctions are suitable as volatile threshold switching elements [5-7, 22-24].

Previous reports have focused on a broad range of film qualities, ranging from amorphous [2–4] to epitaxial quality [8, 9]. Films grown on $Al_2O_3$ (0001) substrates using magnetron reactive sputtering resulted in $NbO_2$ (110) and (111) crystal orientations [8], while films grown on (La,Sr)$_2$(Al,Ta)$_2$O$_6$ (111) substrates using molecular beam epitaxy had both (100) and (320) out-of-plane orientations [9].

In this work we demonstrate that control of the stoichiometry and defect density is possible for the growth of niobium oxide films, with a gradual transformation of the thin film phases from NbO$_2$ to Nb$_2$O$_5$, while maintaining the epitaxial quality. Transport measurements were performed in the current-in-plane geometry using top interdigitated electrodes (IDEs), that is, no bottom electrode was needed, which eliminates the risk of jeopardizing the film quality.

## 2. Experimental

NbO$_2$ films were grown on pre-polished Al$_2$O$_3$ (0001) substrates by pulsed laser deposition from a ceramic Nb$_2$O$_5$ target. The target was prepared from 99.99% Nb$_2$O$_5$ powder (Sigma-Aldrich) pressed into a pellet and sintered for 72 h in air at 1300C. The distance between target and substrate was 7.3 cm. The KrF laser energy density at the target was approximately 2 J/cm$^2$ and its pulse repetition rate was 5Hz. Films were grown at a 650 °C substrate temperature in an O$_2$/Ar 7%/93% gas mixture atmosphere with the total pressure ranging between 1 and 20 mTorr. The surface quality of the substrates and films was monitored using *in situ* Reflection High-Energy Electron Diffraction (RHEED). A four-axis goniometer x-ray diffraction (XRD) system with a Rigaku Cu K$_\alpha$ rotating anode and a Huber goniometer was used for structural characterization of the deposited films. The in-plane epitaxial relationship between the film and the substrate was established by x-ray $\Phi$-scans. Rocking curves were measured using Bruker D8 Discovery X-ray diffractometer. In order to minimize the effect of strain, all films had the same total thickness of 20 nm. Information about the surface oxidation states was obtained using x-ray photoelectron spectroscopy (XPS). The XPS data were calibrated with respect to the carbon 1s peak at 284.8 eV. Film thickness and surface roughness analyses were performed by x-ray reflectivity (XRR) using the x-ray diffraction system mentioned above and by atomic force microscopy (AFM). Optical reflectance and transmittance were measured using a Fourier-transform Infrared (FTIR) spectrometer with a halogen light source and a liquid nitrogen cooled HgCdTe detector. Spectra were obtained for photon energies between 0.347 eV and 1.451 eV with 0.001 eV resolution. Absorption was calculated from experimental data with some offset corrections due to losses during light collection for reflectance measurements.

IDEs were composed of a 50 nm Pt film grown via sputtering at room temperature and patterned on top of the film via the standard photolithography. The IDEs had 25 fingers from each side with a length of 500 μm and a width and finger gap of 5 μm. The voltage was ramped at a rate of 0.2 V/s with a step size of 0.1 V. For pulsed-field measurements triangular ramp voltage pulses were applied to a thin film sample with an attached serial test resistance of $R_s=50\Omega$. Each pulse had a peak amplitude of 20V and time period ($\tau$) ranging from 10s to 1ms. The leakage current was read from the voltage drop across the test resistance.

### 3. Results and discussion

Figure 1(a) shows a typical streaky RHEED pattern observed during and after the film growth, indicating that the film surface was relatively smooth. The RHEED spacing between streaks for $Al_2O_3$ substrates and films grown in 1-15 mTorr was practically the same, while the films grown in 20 mTorr had a smaller RHEED pattern spacing. As discussed in more detail below, the large and small spacings correspond to the formation of $NbO_2$ and $Nb_2O_5$, respectively. The pattern of the films grown in 1-15 mTorr also had six-fold azimuthal rotational anisotropy, in agreement with the existence of three structural twin domains measured via x-ray diffraction, as discussed below.

Thin film interference fringes around the main film peaks in XRD θ-2θ scans (Fig. 1), originating from interference between x-rays reflected from the top and bottom surfaces of the films, indicated relatively low film roughness. All films grown in the pressure range of 1 - 15 mTorr had single (110) out-of-plane orientation in terms of the low temperature distorted rutile unit cell (Fig. 1), in agreement with previous results [8]. The growth can be best understood (Fig. 1c) if the rutile subcell ($a_R$= 4.844 Å and $c_R$=2.993 Å [16]) is considered instead of the conventional tetragonal supercell ($a_T$ and $c_T$), with $[100]_R \parallel [110]_T$ and $[001]_R \parallel [001]_T$. Using the rutile subcell, the $(110)_T=(100)_R$ $NbO_2$ growth is similar to that of (100) $VO_2$ on $Al_2O_3$ (0001) substrates, if the high temperature rutile unit cell of $VO_2$ is considered (equivalent to the (010) low temperature $VO_2$ phase [8]).

The films grown in 20 mTorr pressure had a main peak, indicated by the arrow in Fig. 1b, which could not be assigned to $NbO_2$ and should correspond to $Nb_2O_5$. Due to the pronounced polymorphism of $Nb_2O_5$ [25], the exact estimation of the corresponding crystal $Nb_2O_5$ phase is difficult and is beyond the scope of this paper. Assuming that the initial layer growth took place in the form of $NbO_2$ with the subsequent post-oxidation to $Nb_2O_5$ due to the favorable lattice match with the substrate, TT-, T- and B-$Nb_2O_5$ phases [25] could be successively or simultaneously formed at the substrate temperature of 650 °C, with the corresponding unit cells of a=3.607 Å and c=3.925Å (pseudo-hexagonal [26]), a=6.168 Å, b=29.312 Å, c=3.936Å (orthorhombic L-phase [27]), and a=12.744 Å, b=4.885Å, c=5.563Å, β=105.03° (monoclinic [27]).

The main peak in Fig. 1b for the 20 mTorr sample can then be ascribed to (101), (181) or (020) reflections of TT, T or B-$Nb_2O_5$, respectively. The smaller peak next to it, indicated as "#" in Fig. 1b, can thus belong either to the strained (110) $NbO_2$ or to the (40-2) B-$Nb_2O_5$. A gradual appearance of $Nb_2O_5$ phase with increasing $O_2$ / Ar pressure was observed in the form of an increasing asymmetry around the (110) $NbO_2$ peaks. The $Nb_2O_5$ phase is predominant in the film grown in 20mTorr, as verified by XPS, optical absorption and I-V measurements discussed below. Evidently, increasing $O_2$ partial pressure during the growth resulted in increasing oxidation rates for $Nb^{5+}$ vs. $Nb^{4+}$ ablated from the $Nb_2O_5$ ceramic target, while the additional Ar content helped to establish a relatively slow thin film growth.

Rocking curve scans (Fig. 2a) were performed around the (440) and (0006) reflection of $NbO_2$ film grown in 10 mTorr total pressure and $Al_2O_3$ substrate, and the corresponding full width at half maximum (FWHM) values were found to be $0.02°$ and $0.003°$, respectively, thus indicating the high degree of the out-of-plane and in-plane crystalline orders in $NbO_2$ thin films. Our FWHM values are also lower than those reported in Ref. 9 (FWHM=$0.07°$) and those in Ref. 8 (FWHM=$0.18°$). The in-plane orientation of the film with respect to the substrate was determined from XRD $\Phi$-scans of the (202) and (400) peaks for the $Al_2O_3$ substrate and $NbO_2$ films, respectively. The right inset in Fig. 2a shows results for the film grown in 10 mTorr pressure. A six-fold rotational symmetry of the film was found, which is explained by the presence of three twin $NbO_2$ growth domains with the corresponding {001} axis of the tetragonal $NbO_2$ ∥ {1 -1.0} axis of the $Al_2O_3$ substrate (see also Fig. 1c), i.e. along the oxygen sub-lattice main axis (in agreement with Ref. 8), thus confirming the in-plane epitaxy of the films.

XRR scans shown in Fig. 2b confirmed that the thickness of all samples was approximately 20 nm, 19.4 ± 0.6 nm, and the surface roughness obtained from the XRR fits was approximately 0.5 nm. The atomic force microscopy images (left inset, Fig. 2a for the $NbO_2$ film grown in 1 mTorr) showed that the surface was smooth with root mean square roughness of 0.3 nm, in good agreement with the XRR data.

XPS Nb 3d level core spectra are shown in Fig. 3. Films grown in 1 to 15 mTorr exhibited two $3d_{5/2}$ level peaks of Nb at 205.4 and 206.9 eV. With decreasing growth pressure, the intensity of the peak at 205.4eV increased while the intensity of the peak at 206.9 eV decreased. Thus, the 205.4 eV and 206.9 eV peaks are related to lower and higher Nb valencies, respectively. The reference spectrum taken on $Nb_2O_5$ powder (solid line curve in Fig. 3) confirmed that the most intense peak at 206.9 eV was from $3d_{5/2}$ level of $Nb^{+5}$. Similarly, $3d_{3/2}$ level peaks were found at 209.6 eV and at 208.2 eV for $Nb^{+5}$ and $Nb^{+4}$, respectively, as shown by de-convoluted peaks in the inset to Fig. 3a. There are two views about XPS $NbO_2$ data in the literature. One group of authors [7, 10, 28] claims that the peaks at 206.9 eV and at 205.4-205.6 eV correspond to $3d_{5/2}$ peaks of $Nb^{5+}$ and $Nb^{+4}$, respectively, while another group [9, 29-31] assigns them to $Nb^{+4}$ and $Nb^{3+}$. Our interpretation agrees with that of the first group. Therefore, decreasing the total growth pressure with constant oxygen mass flow content helped to reduce the nominal $Nb^{5+}$ in $Nb_2O_5$ to $Nb^{4+}$. The presence of the XPS peaks identified as $Nb^{5+}$ for the samples which were characterized by XRD as pure $NbO_2$ (e.g. sample grown at 1 mTorr) can be explained by the existence of a thin (1-2 nm) surface $Nb_2O_5$ layer formed after exposure to atmosphere [10, 11].

Based on the structural characterization, we identified samples grown in 1 mTorr and 20

mTorr as $NbO_2$ and $Nb_2O_5$ phases, respectively. This was confirmed by optical band gap measurements shown in Fig. 3b. The sample grown at 1 mTorr exhibited a band gap of approximately 0.6 eV, in agreement with the previous literature values reported 0.3 - 0.4eV [10], 0.5 eV [32], 0.7 eV [11], 0.88 eV [33], and at least 1.0 eV [9]. The sample grown in 20 mTorr pressure showed no significant absorption within the photon energy range, in agreement with the higher $Nb_2O_5$ band gap energy ranging from 3.5 to 4.8 eV [11, 34, 35]. The small, broad peaks near 0.65 eV and 0.9 eV are due to water absorption in air. In order to further determine the nature of the band gap in $NbO_2$, Tauc plots were graphed (inset of Fig. 3b), i.e. the absorption coefficient α was plotted as $(\alpha\hbar\omega)^n$ vs. the photon energy $\hbar\omega$ with n=1/2 and n=2, which are related to indirect and direct transitions, respectively [36]. The best proportionality was found for n=1/2, thus confirming an indirect band gap of 0.57 eV for $NbO_2$, in agreement with band structure calculations by Weibin et al. [11].

Current vs. voltage (I-V) characteristics of $NbO_2$ lateral devices were measured with the electric field applied in-plane using IDEs (Fig. 4a). Typical threshold-switching behavior was observed, which is related to a reversible MIT driven by the local heating effect. The threshold-electric field magnitude was found to be $E_{th}$=2.2 X $10^6$ V/m. Typical threshold fields reported for vertical devices are on the order of $10^8$ V/m [5, 7]. Thus, a purely electric field-induced effect is unlikely to be the cause of the observed switching.

The film grown at 20 mTorr showed a different behavior (Fig. 4b), in agreement with its predominant $Nb_2O_5$ nature. This film had a much larger resistance than the $NbO_2$ films so that an electric field above 34.6 x $10^6$ V/m was needed to switch the initial low current state to a high current state, which remained after the electric field was removed. During the reversed cycle a transition to further higher current state occurred, that is, the switching behavior was irreversible. No $Nb^{+4}$ valencies were found in this film by the surface sensitive XPS technique (Fig. 3a). The most likely switching scenario for the film grown in 20 mTorr pressure involves oxygen migration and electroforming processes typical for $Nb_2O_5$ systems [6].

Fig. 4c shows the corresponding variation of the temperature during the measurements of I-V cycles on the $NbO_2$ film grown in 15 mTorr pressure. The temperature was read out by the thermocouple temperature sensor located on the copper block on which the sample was mounted and in close proximity to the sample, so the temperature readings provide a remote measure of the dissipated power in the thin film. Three regions A, B, and C can be identified in the corresponding temperature vs. voltage curve, related to the current flow while being in the high resistance state, to the increased current after the switching into the low resistance state, and to the temperature increase while the current remained constant (due to the instrumental compliance setting), respectively. The temperature in region B followed the abrupt increase in the current curve due to the Joule heating in $NbO_2$ thin films. A continuous decrease in temperature was observed until the voltage reached 5 V (inset, Fig. 4c), most likely due to the Peltier effect at the metal-semiconductor junction [37, 38]. It is important to note that the temperature increase observed in Fig. 4c does not represent the real temperature in the film during the switching. The first reason is that our experimental geometry was not suitable for sensing the local heating effect of the film. Rather, the maximal temperature rise to 343 K is indicative for the thermal gradient between the film and the attached copper block, separated by a 0.5mm thick $Al_2O_3$ substrate promoting the heat transfer. The second reason is that, as discussed below, we estimate a relatively small fraction of the total film volume where MIT threshold switching takes place, and therefore the average film temperature was likely below the MIT transition temperature for bulk $NbO_2$ (1081 K).

In contrast to the previous reports [3-7] where non-connected conducting filaments pre-existed and/or were modified by electroforming in non-stoichiometric and amorphous $NbO_x$ phases, we observed volatile and reversible switching in the epitaxial quality $NbO_2$ films with minimal non-stoichiometry and with no electroforming steps required. Therefore we assume that threshold switching in our devices follows conduction paths associated with defects such as twin domain boundaries, in which thermally confined filaments are locally heated above the MIT temperature [39, 40, 41]. Alternatively, filamentary paths in $NbO_2$ films could be confined to the interface with the surface $Nb_2O_5$ layer similar to what was observed in Ref. 39. Note that a recent study of nanodevices made from $NbO_2$ found no significant contribution to transport properties of $NbO_2$ films due to the presence of ~2 nm top surface insulating layer of $Nb_2O_5$ [42]. Assuming that the current flows through the entire $NbO_2$ film volume located between the top IDE fingers, the dissipation power per volume imparted at the threshold switching voltage of the I-V curves (Fig. 4a) can be estimated to be $1.3 \times 10^{-4}$ nW/nm$^3$, which is much less than the 5 nW/nm$^3$ reported for 20 nm thick and 60nm in diameter electroformed $NbO_2$ channel vertically stacked between 110 × 110 nm$^2$ Pt contacts [1, 2]. Despite certain differences in the film crystalline quality between those devices (e.g. density of defects), it appears more likely that the discrepancy between power per volume ratios is due to the limited small fraction of the total volume of $NbO_2$ conducting channels (filaments) in our lateral devices, where the MIT-triggered threshold switching takes place, similar to the reports in Refs. 1, 2 and 39.

Time dependence of the switching was studied using triangular voltage pulses (Fig. 5b).

The principal connection scheme is represented in Fig. 5a. We observed a non-linear current response (Fig. 5c), corresponding to the threshold switching behavior, for the time constant $\tau >$ 1 ms. At the same time, the maximal current flowing at the half pulse period decreased with decreasing $\tau$ values, that is, faster voltage pulses initiate increasingly linear current responses. In addition, non-linearity in the current response started to appear at lower voltage values when voltage pulses with longer time periods $\tau$ (i.e. with lower voltage ramp rates) were applied. In total, this suggests that the nature of the threshold current switching in $NbO_2$ thin films is due to Joule self-heating rather than an electric field-induced effect, in agreement with I-V characteristics shown in Fig. 4a, where no abrupt transition was observed. The threshold switching in our devices is thus much slower than those reported in Ref. 2, where characteristic ON-OFF and OFF-ON times were of the order of nanoseconds. A larger gap between the electrodes (5 μm vs. 20 nm in Ref. 2) and the complexity of electric field lines in our case (IDE electrodes vs. top and bottom Pt electrodes used in Ref. 2) could explain this discrepancy.

## 4. Conclusions

We have demonstrated the growth of single phase epitaxial $NbO_2$ thin films on $Al_2O_3$ (0001) substrates. The oxidation state of Nb was controlled by the total growth pressure of $O_2$/Ar mixture. XRD and XPS analysis confirmed the formation a pure phase $NbO_2$ and major phase $Nb_2O_5$ in the films grown in 1 and 20 mTorr total growth pressure, respectively. I-V measurements of two terminal lateral devices showed reversible threshold resistance switching behavior with switching fields of approximately 2.2 x $10^6$ V/m and 34.6 x $10^6$ V/m for films with predominant $NbO_2$ and $Nb_2O_5$ phases, respectively. Our results demonstrate that the stoichiometric and epitaxial $NbO_2$ thin films show volatile threshold resistance switching. $Nb_2O_5$

films exhibited irreversible phase transformation when relatively large electric fields were applied. Pulsed field-measurements yielded typical switching times above 1ms in lateral devices made of $NbO_2$.

**Acknowledgements**

This work was supported in part by FAME, one of six centers of STARnet, a Semiconductor Research Corporation program sponsored by MARCO and DARPA (Contract # 2013-MA-2382), a Research Challenge Grant from the WV Higher Education Policy Commission (HEPC.dsr.12.29), and the WVU Shared Research Facilities.

Figures

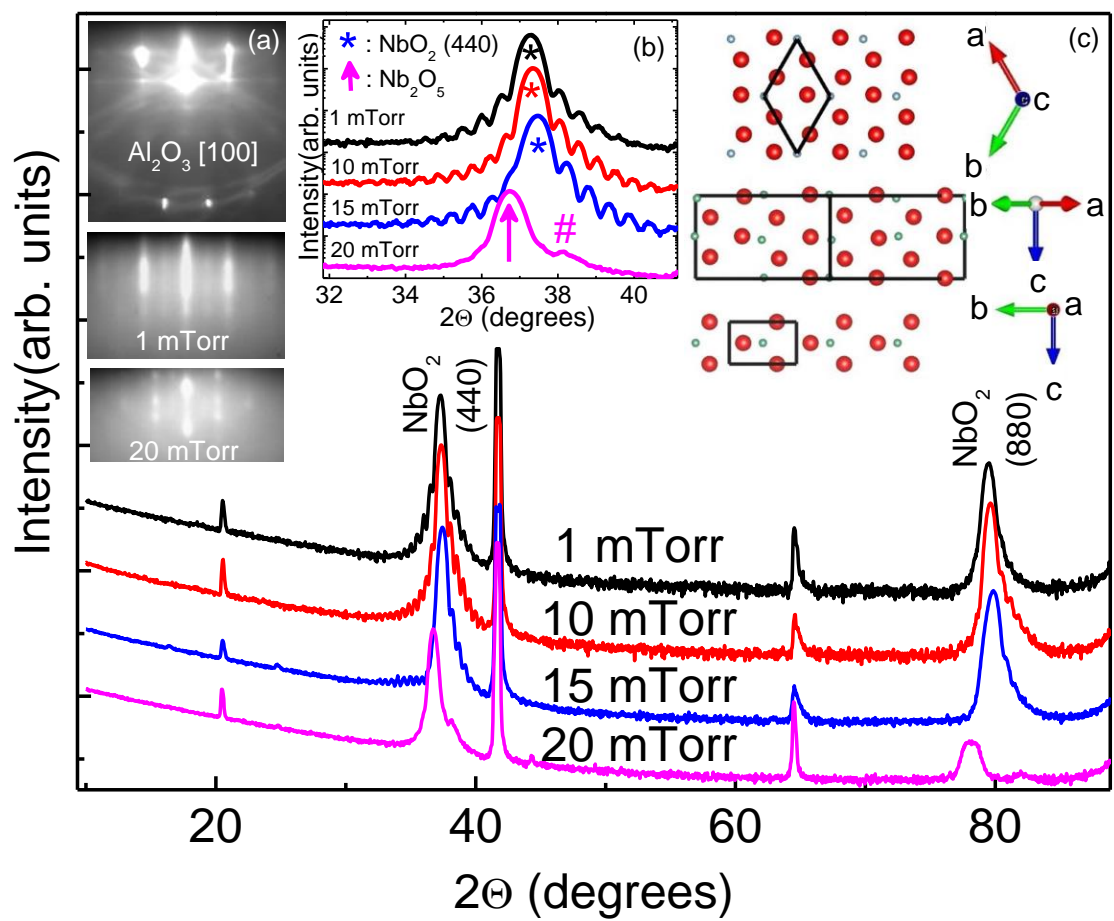

Fig. 1: X-ray diffraction spectra from samples grown at different growth pressure. Insets: (a) same-scale RHEED images of the substrate (top) and the films grown in 1 mTorr (middle) and 20mTorr (bottom) (b) extended view of NbO$_2$ (440) [stars] and Nb$_2$O$_5$ [arrow] peaks measured on films grown in 1-15 mTorr and 20 mTorr pressure, respectively. (c) Same-scale top view of the substrate and film surfaces, mutually oriented in the same way as in the grown film samples. From the top to the bottom: (00.1) Al$_2$O$_3$, low temperature tetragonal (110) NbO$_2$, high temperature rutile (100) NbO$_2$. Corresponding lattice axes are shown to the right. Solid black lines signify unit cells. Large and small spheres denote anions and cations, respectively.

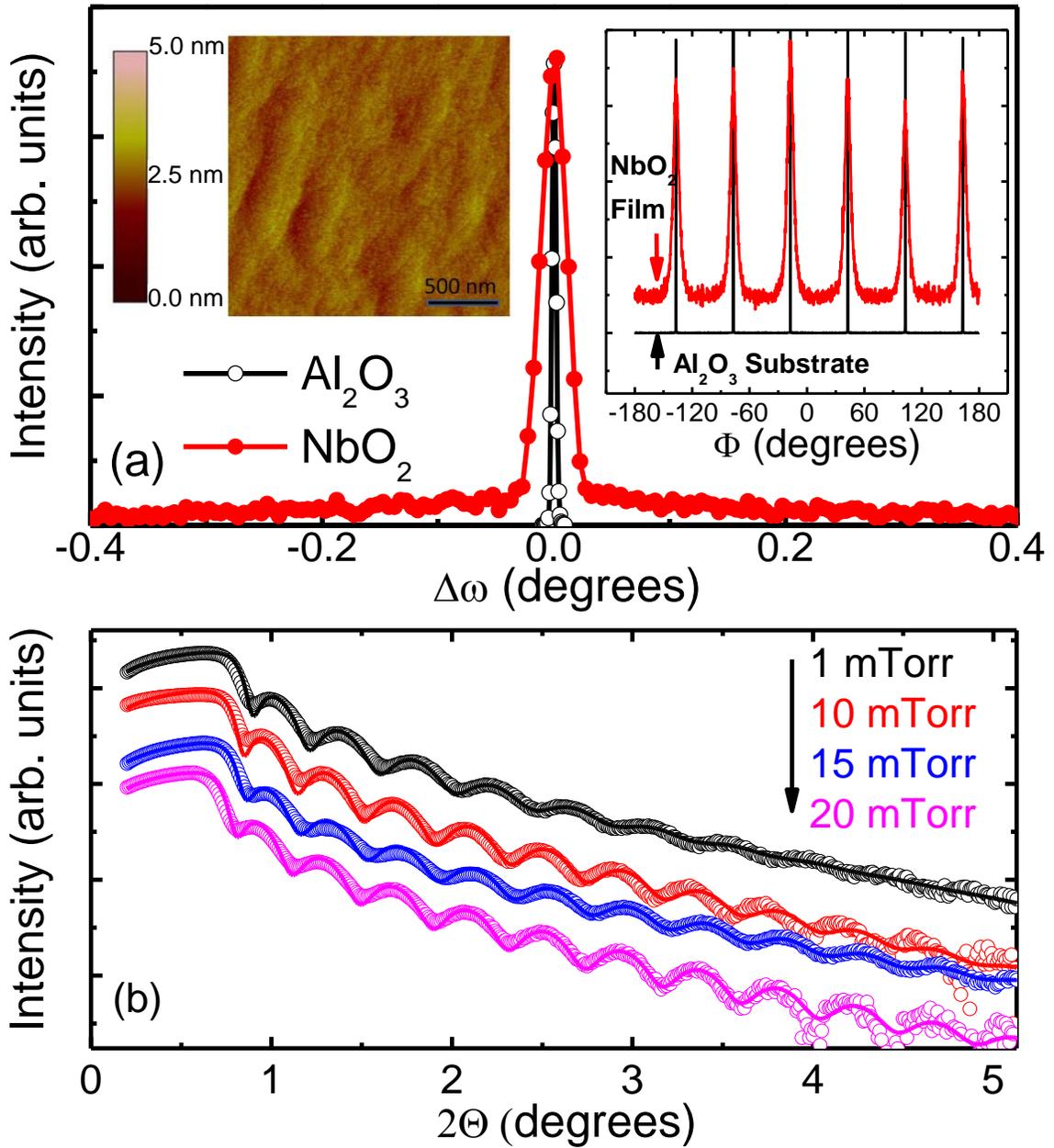

Fig. 2: (a) Rocking curve of (440) and (0006) peak of NbO$_2$ film grown in 10 mTorr pressure and Al$_2$O$_3$ substrate, respectively. Left inset: atomic force microscopy image of the NbO$_2$ surface for the film grown in 1 mTorr. Right inset: $\Phi$-scans of (202) and (400) peaks from the Al$_2$O$_3$ substrate and NbO$_2$ film grown in 10 mTorr, respectively. (b) X-ray reflectivity spectra measured

on NbO$_2$ films grown in different pressures (open circles) with the corresponding fits (solid lines).

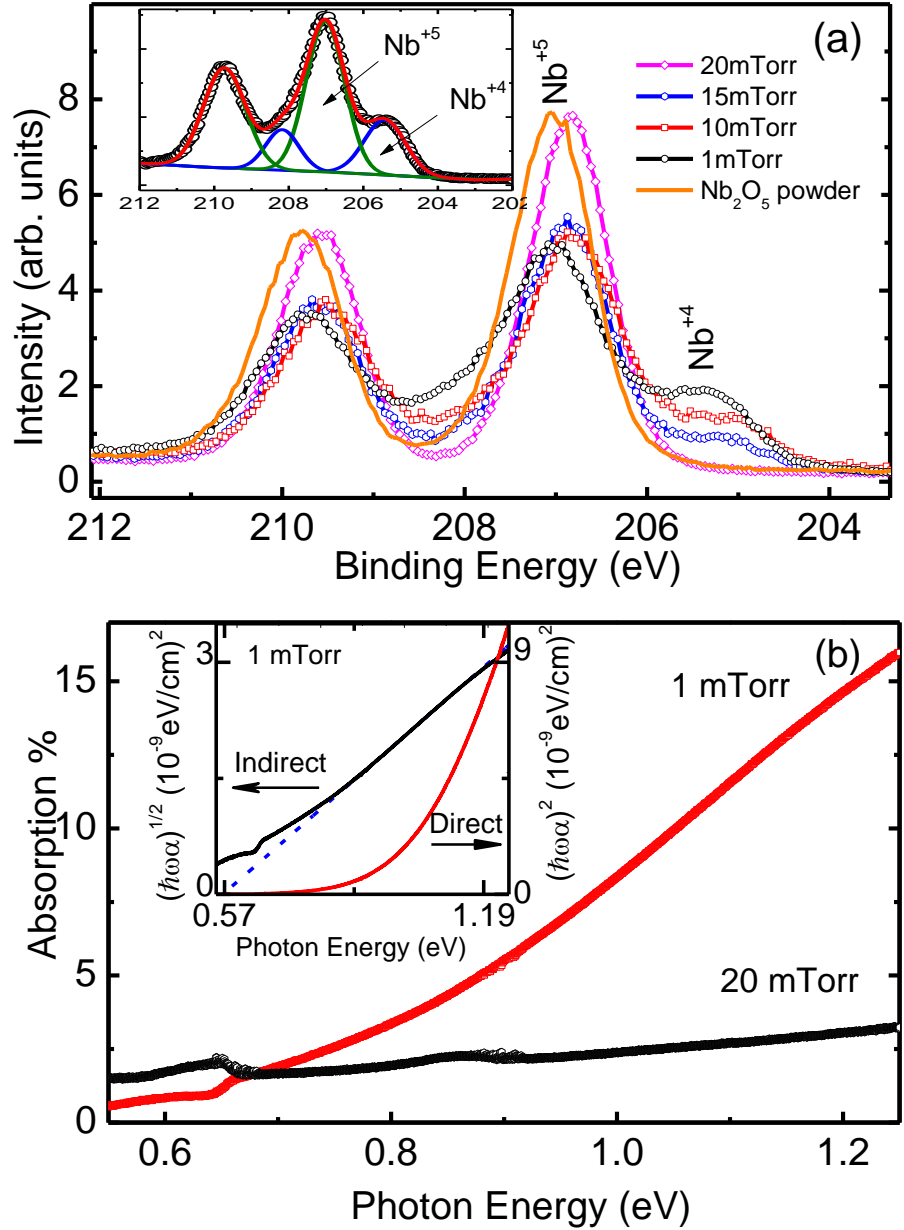

Fig. 3: (a) X-ray photoelectron spectroscopy pattern of the NbO$_2$ films grown in different growth pressures. Inset shows de-convoluted peaks for the film grown in 10 mTorr. (b) Optical absorption spectra for the predominant NbO$_2$ and Nb$_2$O$_5$ film phases grown in 1 and 20 mTorr, respectively. The inset shows Tauc plots, $(\alpha\hbar\omega)^n$ vs. $\hbar\omega$, where $\alpha$ is the absorbance and $\hbar\omega$ the photon energy, for the NbO$_2$ film grown in 1mTorr. Left and right scales correspond to indirect (n=1/2) and direct (n=2) optical transitions, respectively. Blue dashed line signify linear fit for the indirect band gap E$_g$=0.57 eV.

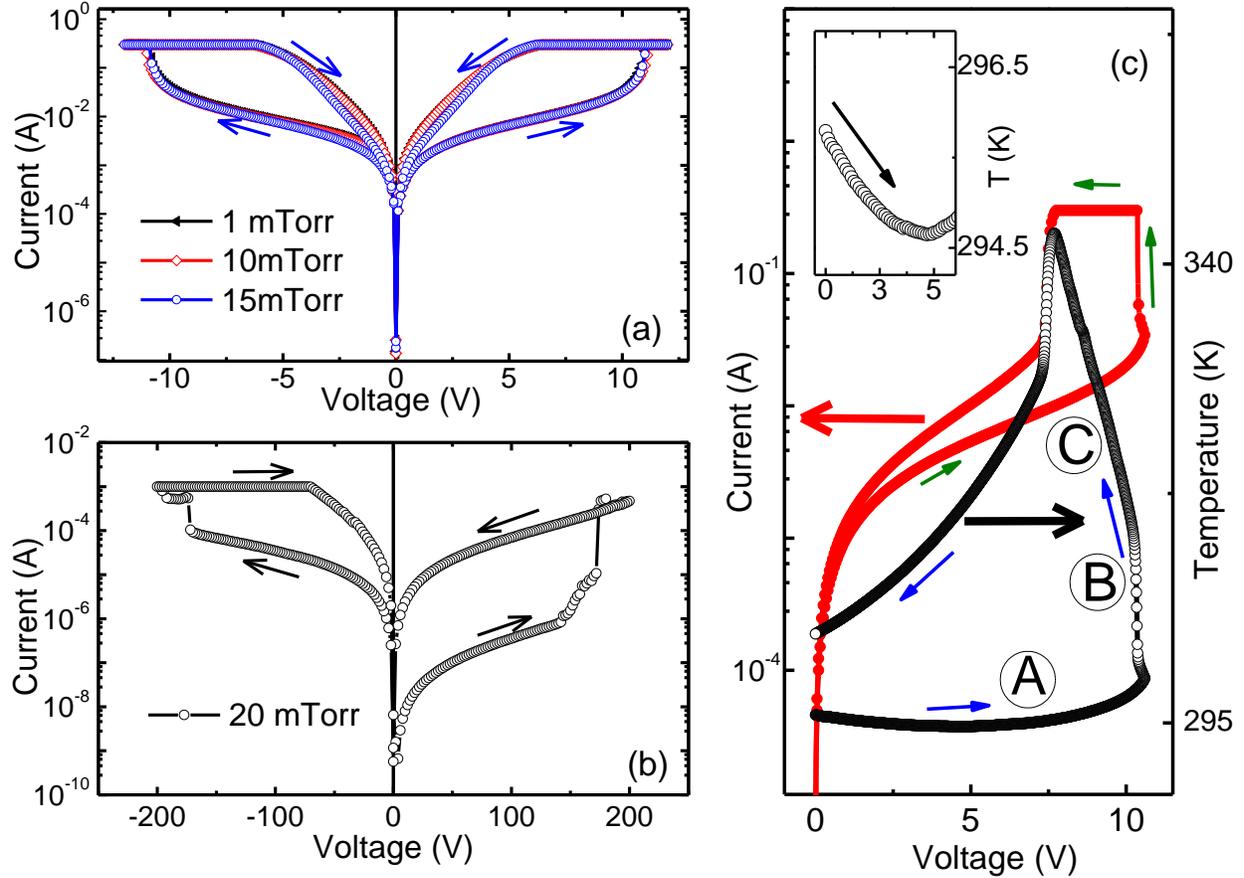

Fig. 4: (a) I-V characteristics measured on the films grown in 1, 10 and 15 mTorr. (b) I-V characteristics measured on the film grown in 20 mTorr. (c) I-V curve (red filled circles) and the simultaneous temperature variation (black open circles) measured on the film grown in 15 mTorr. The flat regions in the current are due to compliance limits of 0.3 A in (a), 10$^{-3}$ A in (b) and 0.25 A in (c).

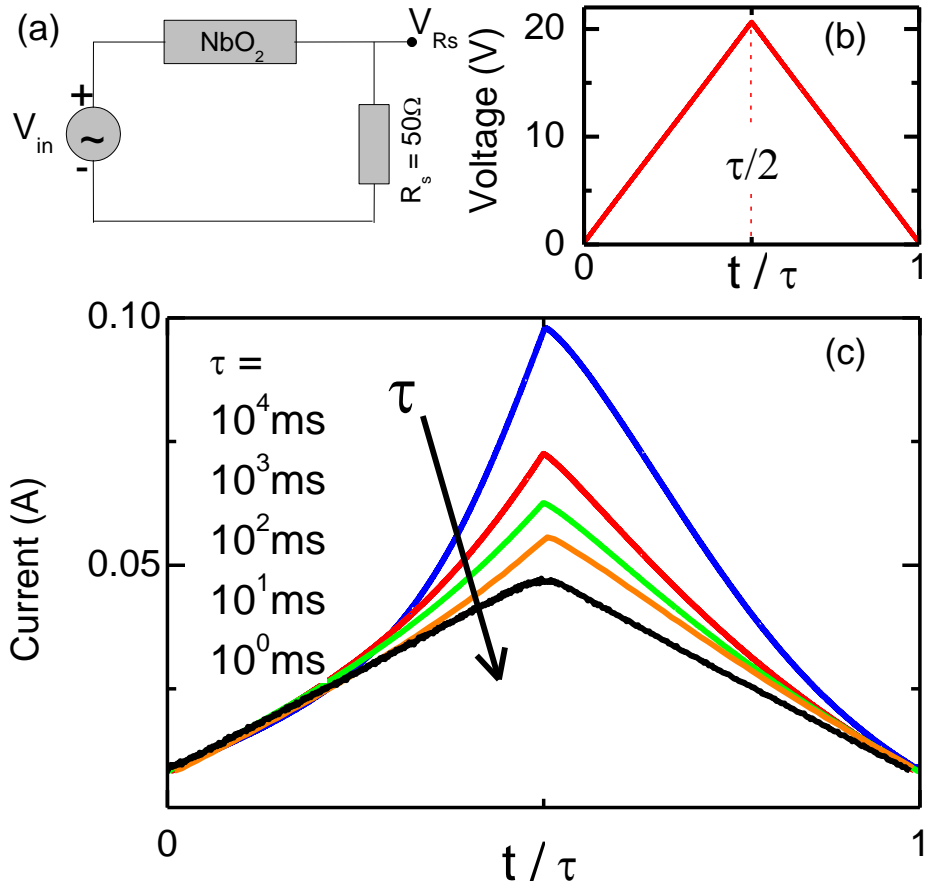

Fig. 5 (a) principal electric scheme of the pulsed-field measurements. (b) and (c) show applied voltage and measured current pulses vs time in relative units of time periods τ, respectively.